\documentclass[showpacs, twocolumn,prb]{revtex4-1}
\usepackage{epsfig}
\usepackage{amssymb,amsmath,amsthm,graphics}
\usepackage{color}
\usepackage[colorlinks=true,linkcolor=blue,citecolor=blue]{hyperref}

\begin{document}

\title{Spin-transfer driven nano-oscillators are equivalent to parametric resonators}

\author{Alejandro O. Le\'on} 
\email{aoleon@dfi.uchile.cl}
\author{Marcel G. Clerc}
\email{marcel@dfi.uchile.cl} 
\affiliation{Departamento de F\'{i}sica, Facultad de Ciencias F\'isicas y Matem\'aticas,
Universidad de Chile, Casilla 487-3, Santiago, Chile.}

\begin{abstract}
The equivalence between different physical systems permits to transfer 
knowledge between them and to characterize the universal nature of their dynamics. 
We demonstrate that a nanopillar driven by a spin-transfer torque is equivalent to a rotating magnetic plate, which permits to consider the nanopillar as a 
macroscopic system under a time modulated injection of energy, that is, a simple parametric resonator. 
This equivalence allows us to characterize the phases diagram and to predict magnetic states 
and dynamical behaviors, such as solitons, stationary textures and oscillatory localized states, 
among others. Numerical simulations confirm these predictions.
\end{abstract}

\pacs{
05.45.Yv, 
89.75.Kd 
75.78.-n  
}

\maketitle 
\section{Introduction}
\begin{figure}[b!]
\includegraphics[width=8.6 cm]{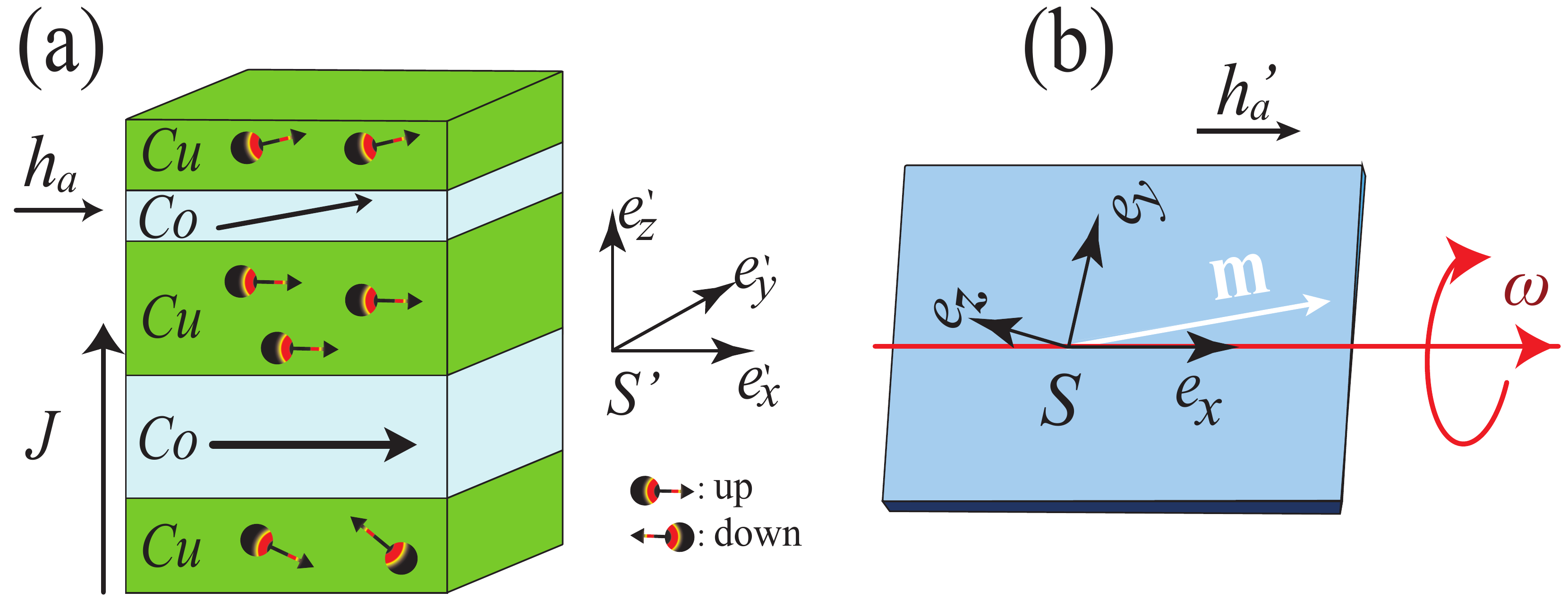}
\caption{(color online) Equivalent physical systems. (a) Schematic 
representation of the spin-transfer torque nano-oscillator setup.
The light (blue) and dark (green) layers represent magnetic and non-magnetic 
metal films, respectively.  
$J$ and $h_a$ are the electric current through the spin-valve and the external 
magnetic field, 
both effects are parallel to the easy axes of the ferromagnetic layer under study. 
${\bf M}_o$ stands for the magnetization of the fixed layer. 
 (b)  Rotating magnetic plate with an easy axis  
in the rotation direction, subjected to a constant magnetic field, ${\bf h'_a}$.} 
\label{fig-Nanopillar}
\end{figure}

Current-driven magnetization dynamics have attracted much attention in recent years, 
because of both the rich phenomenology that emerges and the promising 
applications in memory technology~\cite{Maekawa}. 
A remarkable example occurs when a direct spin-polarized current applies a torque to 
nanoscale ferromagnets, effect known as spin-transfer torque~\cite{Slon,Berger-ST}. This effect has 
been confirmed experimentally  \cite{Tsoi,Sun99,Ralph1999,Ralph2000,ST_EXP,ST_EXPIII}, 
particularly, observation of magnetization reversal caused by Spin-transfer torques 
was reported in \cite{Ralph1999,Ralph2000,ST_EXPI,Berkov}.
Spin-transfer effects are usually studied in the metallic multi-layer
nanopillar, or spin-valve, depicted in Figure~\ref{fig-Nanopillar}a, where two magnetic films (light layers), 
the {\it free} and the {\it fixed} one, are separated by a non-magnetic spacer (darker layer).
In such nanopillar, an electric current  $J$ applied through the spin-valve  
transfers spin angular momentum from the film with {\it fixed} magnetization 
to the {\it free} ferromagnetic layer. 

When the direct current overcomes a critical value, the spin-transfer torque destabilizes 
the state in which both magnetizations point parallel, and the free magnetization 
switches or precesses in the microwave-frequency domain. Most scientific efforts 
have focused on this regime, in which the free magnetization behaves as an self-oscillator 
with negative damping~\cite{Slavin}. Another interesting case is when there is an external 
field that disfavors the parallel state and the spin-polarized current favors it, under this 
regime, it is expected that the system will generate complex dynamics 
as a result of both opposing effects.

The aim of this article is to show that nanopillars under the effect of a spin-polarized 
direct electric current exhibit the same dynamics present in systems 
with a time modulated injection of energy, known as Parametric systems \cite{Landau}. 
Parametric systems oscillate at the half of the forcing frequency, phenomenon known as parametric resonance. Examples of parametric systems are a layer of water oscillating vertically\cite{WaterSoliton}, localized structures in nonlinear lattices
\cite{NonLinearLatices}, light pulses in optical fibers \cite{OpticalFiber},  optical parametric
oscillators \cite{OPO}, easy-plane ferromagnetic materials exposed to an oscillatory magnetic field \cite{Barashenkov}, to mention a few. 

To understand the  parametric nature of the
spin-transfer-driven nanopillars, we put in evidence that this system is equivalent to a simple 
rotating magnetic plate subjected to a constant magnetic field applied in the rotation 
direction (see Fig.~\ref{fig-Nanopillar}b).
Where the electric current intensity on the nanopillar corresponds to the angular 
velocity in the equivalent rotational system. 
We analytically show that the magnetization dynamic 
of  a nanopillar under the effect of a spin-transfer torque
is well described by {\it the  Parametrically Driven, damped NonLinear Schr\"odinger equation} (PDNLS). 
This equation is the paradigmatic model of parametric systems with small injection and dissipation
of energy \cite{ClercCoulibalyLaroze2008}.
Based on this model we predict that the spin-transfer torque generates equilibria, solitons, oscillons, 
patterns, propagative walls between symmetric periodic structures, 
complex behaviours, among others. Numerical simulations of the Landau-Lifshitz-Gilbert equation 
confirm these theoretical predictions. 

The manuscript is organized as follows, in the next section we present the nanopillar and the equation of motion of an homogeneous free magnetization. In Sec.~\ref{Sec-Equivalence}, we analyze the relation between the nanopillar and parametric systems.  In Sec.~\ref{Sec-Inhomogenity} we explore the inhomogeneous dynamics predicted by the parametric nature of the Spin-transfer torque effect at dominant order. Finally in Sec.~\ref{Sec-Conclusions} we give the conclusions and remarks.

\section{Macrospin dynamics of the free layer}
\label{Sec-MagDyn}
Consider a nanopillar 
device, with fixed layer magnetization $\mathbf{M}_0$ along the positive $x$-axis 
as depicted by Fig.~\ref{fig-Nanopillar}, this ferromagnet has a large magnetocrystalline 
anisotropy or it is thicker than the free layer, and therefore it acts as a polarizer for the electric current. 
Let us assume that the free layer is a single-domain magnet, it is, the magnetization rotates uniformly 
$\mathbf{m}(\mathbf{r},t)=\mathbf{m}(t)$.

Hereafter, we work with the following adimensionalization, 
the magnetization of the free layer $\mathbf{M}\rightarrow M_s \mathbf{m}$ and the external field $\mathbf{H_a}\rightarrow M_s \mathbf{h_a}$ are normalized 
by the saturation magnetization $M_s$; the time $t\rightarrow \gamma M_st$ 
is written in terms of the gyromagnetic constant $\gamma$, and $M_s$. 
For instance, in a Cobalt layer of $3nm$ of thickness, $M_s\simeq 1.4$ $10^6 A/m$, 
and the characteristic time scale is $(\gamma M_s)^{-1}\simeq 3.2 ps$~\cite{Serpico}.

When the free magnetization is homogeneous, the normalized magnetic energy 
per unit of volume is\cite{Serpico}
\begin{eqnarray}
\frac{E}{\mu_0 M^2_s}=-\mathbf{m}\cdot {\bf h_a}
- \frac{1}{2}\beta_xm_x^2
+ \frac{1}{2}\beta_z m_z^2,
\label{Eq-Energy}
\end{eqnarray}
the external magnetic field ${\bf h_a}=h_a \mathbf{e}_{x}$ points along the 
$x$-axis (see Fig.~\ref{fig-Nanopillar}). 
The coefficients $\beta_x$ and $\beta_z$ 
are combinations of the normalized anisotropy and demagnetization 
constants with respect to the appropriate axes, where $\beta_x$ ($\beta_z$) 
favors (disfavors)  the free magnetization  in the $x$-axis ($z$-axis).

The dynamic of the magnetization of the free layer is described by the dimensionless Landau-Lifshitz-Gilbert
equation (LLG) with an extra term that accounts for the spin-transfer 
torque~\cite{Serpico,Slon,Ralph1999,Ralph2000,ST_THEORY}

\begin{equation}
\frac{d \mathbf{m}}{d t}=-\mathbf{m}
\times \mathbf{h}_{eff}
+\alpha\mathbf{m}\times\frac{d \mathbf{m}}{d t}
+g\;\mathbf{m}\times(\mathbf{m}\times \mathbf{e}_{\mathbf{x}}).
\label{Eq-LLG}
\end{equation}
The first term of the right hand side of Eq.~(\ref{Eq-LLG}) accounts 
for the conservative precessions generated by the effective field,
\begin{equation}
\mathbf{h}_{eff}\equiv -\frac{1}{\mu_0M_s^2}\frac{\delta E}
{\delta\mathbf{m}}=(h_{a}+\beta_{x}m_{x})\mathbf{e}_{x}-\beta_{z}m_{z}\mathbf{e}_{z}.
\label{Eq-EffF}
\end{equation}
The second and third terms of Eq.~(\ref{Eq-LLG}) are the 
phenomenological Gilbert damping and the Spin-transfer torque respectively. The dimensionless 
prefactor $g$ is given by\cite{Berkov}
$g\equiv \mathcal{P}(m_x)(\hbar/2)(J/d|e|)(1/\mu_0 M_s^2)$,
 and $\mathcal{P}$ describes the electron polarization at the interface between the magnet 
and the spacer, $J$ the current density of electrons, $d$ the thickness of the layer and  $e<0$ 
the electric charge. The current density of electrons $J$ and the parameter $g$ are negative when the electrons flow from the fixed to 
the free layer. There are different expressions for the polarization $\mathcal{P}(m_x)$ in 
the literature~\cite{Slon,Slon2,Xiao1,Xiao2,Fert}. For certain type of 
nanopillars, a better agreement with experimental observations is obtained 
if $\mathcal{P}(m_x)$ is constant, see ref.~\cite{Xiao2,Kim,LeeLee} for more details.

The dynamics of LLG are characterized by the conservation of the magnetization 
magnitude $\| \mathbf{\mathbf{m}}\|=1$,
since $\mathbf{m}$ and $d\mathbf{m}/dt$ are perpendicular.
The LLG model, Eq.~(\ref{Eq-LLG}), admits two natural equilibria $\mathbf{\mathbf{m}}=\pm\mathbf{e}_{x}$,
which represent a free magnetization that is parallel ($+$) or anti-parallel ($-)$ to 
the fixed magnetization ${\bf M}_0$ 
(see Fig.~\ref{fig-Nanopillar}a). Both states correspond to extrema of the free energy $E$. 
We will concentrate on the equilibrium  $\mathbf{\mathbf{m}}=\mathbf{e}_{x}$, 
nevertheless due to the symmetries of the LLG equation, the same results 
hold for $\mathbf{\mathbf{m}}=-\mathbf{e}_{x}$ when replacing $(g,h_a)$ by $(-g,-h_a)$.

\section{Equivalent physical systems} 
\label{Sec-Equivalence}
Let us consider a rotating magnetic 
plane with angular velocity $\mathbf{\Omega}=\Omega_0 \mathbf{e}_x$ 
and  an easy axis  
in the rotation direction, subjected to a constant magnetic field applied in 
the rotation direction ${\bf h_a}'=(h_a+\Omega_0)\mathbf{e}_x$, (see Fig. \ref{fig-Nanopillar}b). 

This rotating ferromagnet can be described in both the co-movil 
frame $S$, defined by the vectors $\{\mathbf{e}_x,\mathbf{e}_y,\mathbf{e}_z\}$, or 
in the inertial frame $S'$, defined by $\{\mathbf{e}_x',\mathbf{e}_y',\mathbf{e}_z'\}$. 
Note that the ferromagnetic easy axis is described by the same vector in the both 
frames $(\mathbf{e}_x'=\mathbf{e}_x)$, nevertheless unit vectors $\mathbf{e}_y(t)
=\cos(\Omega_0 t)\mathbf{e}_y'+\sin(\Omega_0 t) \mathbf{e}_z'$  and $\mathbf{e}_z(t)
=-\sin(\Omega_0 t) \mathbf{e}_y'+\cos(\Omega_0 t)\mathbf{e}_z'$ rotate together 
with the magnetic plate (see  Fig.~\ref{fig-Nanopillar}b). In the co-movil system the 
normalized magnetic energy will be the same of Eq.~(\ref{Eq-Energy}), 
however in the inertial frame the energy depends explicitly in time

\begin{eqnarray}
\frac{E'}{\mu_0 M^2_s}=-\mathbf{m}\cdot {\bf h_a}'
- \frac{1}{2}\beta_xm_x'^2
+ \frac{1}{2}\beta_{zz}'(t) m_z'^2\nonumber\\
+ \frac{1}{2}\beta_{yy}'(t) m_y'^2
+ \frac{1}{2}\beta_{yz}'(t) m_y'm_z',
\label{Eq-EnergyRot}
\end{eqnarray}
where the time varying coefficients $\beta_{zz}'=\beta_z\left(1
+\cos(2\Omega_0 t)\right)/2$, $\beta_{yy}'=\beta_z\left(1-\cos(2\Omega_0 t)\right)/2$, 
and  $\beta_{yz}'=-\beta_z\sin(2\Omega_0 t)$ act as a parametric forcing. 
Note that the frequency of the forcing is twice the frequency of the rotations. Therefore, this system presents a subharmonic parametric resonance\cite{Landau}.

The dynamics of the magnetic plane in the inertial frame $S'$ is described by the 
Landau-Lifshitz-Gilbert equation
\begin{equation}
\left.\frac{d \mathbf{m}}{d t}\right|_{S'}=-\mathbf{m}\times\mathbf{h}_{eff}'(t)+\alpha\mathbf{m}\times  \left. \frac{d \mathbf{m}}{d t}\right|_{S'}.
\label{Eq-LLGRotatingPlateS}
\end{equation}
Where $\mathbf{h}_{eff}'=-(1/\mu_0 M_0^2)(\delta E'/\delta\mathbf{m})$. Let us now write the Eq.~(\ref{Eq-LLGRotatingPlateS}) in the non-inertial frame $S$,
 where the time derivative operator  in the rotating system takes the form $\partial_t|_{S'}=\partial_t|_{S} +\mathbf{\Omega} \times$\cite{Landau}, 
 thus the dynamics of the rotating magnetic plate in the non-inertial frame $S$  reads
\begin{eqnarray}
\left. \frac{d \mathbf{m}}{d t}\right|_{S} =-&\mathbf{m}
\times\mathbf{h}_{eff}+\alpha\mathbf{m}\times \left. \frac{d \mathbf{m}}{d t} \right|_{S}\nonumber \\
&- \alpha  \Omega_0 \mathbf{m}\times (\mathbf{m} \times \mathbf{e}_{x}).
\label{Eq-LLGRotatingPlateS'}
\end{eqnarray}
Where the effective field $\mathbf{h}_{eff}$ is the same of formula (\ref{Eq-EffF}). Therefore, the dynamics of the rotating magnetic plate in the non-inertial frame $S$, Eq.~(\ref{Eq-LLGRotatingPlateS'}), 
is a time independent equation, which is equivalent to the dynamics of a nanopillar under the effect of a 
spin-transfer torque generated by a uniform electric current, Eq.~(\ref{Eq-LLG}). 
In this equivalence, the intensity of spin-transfer effect on the nanopillar $g$ corresponds to the
angular velocity by the dissipation parameter, $-\alpha \Omega_0$. 
Indeed, the two physical systems depicted in Fig.~\ref{fig-Nanopillar} are equivalent.
In the next sections, we will apply the well-known understanding on parametric systems to the nano-oscillator.
 
\subsection{Parametrically Driven damped NonLinear Schr\"odinger equation}
\label{Sec-PDNLS}

To obtain a simple model that permits analytical calculations around the parallel state,
we use the following stereographic representation \cite{Lakshmanan}
\begin{equation}
\psi(\mathbf{r},t)=\frac{m_{y}+im_{z}}{1+m_{x}},
\label{Eq-CofVar}
\end{equation}
where $\psi$ is a complex field. This representation corresponds to consider 
an equatorial plane intersecting the magnetization unit sphere.
The magnetization components are related with the complex field by $m_{x}  = (1-|\psi|^{2})/(1+|\psi|^{2})$, 
$m_{y}  =  (\psi+\bar{\psi})/(1+|\psi|^{2})$ and $m_{z} =  (i\left(\bar{\psi}-\psi\right))/(1+|\psi|^{2})$,
where $\bar{\psi}$ stands for the complex conjugate of $\psi$.
Notice the parallel state  $\mathbf{\mathbf{m}}=\mathbf{e}_{x}$ is mapped to the origin of the $\psi$-plane. 
The LLG, Eq.~(\ref{Eq-LLG}) or Eq.~(\ref{Eq-LLGRotatingPlateS'}), takes the following form 
\begin{eqnarray}
\left(i+\alpha\right)\frac{d\psi}{dt} &=\left(ig-h_{a}\right)\psi-\frac{\beta_{z}}{2}\left(\psi-\overline{\psi}\right)\frac{1+\psi^{2}}{1+|\psi|^{2}} 
\nonumber \\
&-\beta_{x}\psi\frac{1-|\psi|^{2}}{1+|\psi|^{2}}.
\label{Eq-LLGSterographic}
\end{eqnarray}
\begin{figure}[t!]
\includegraphics[width=8.0 cm]{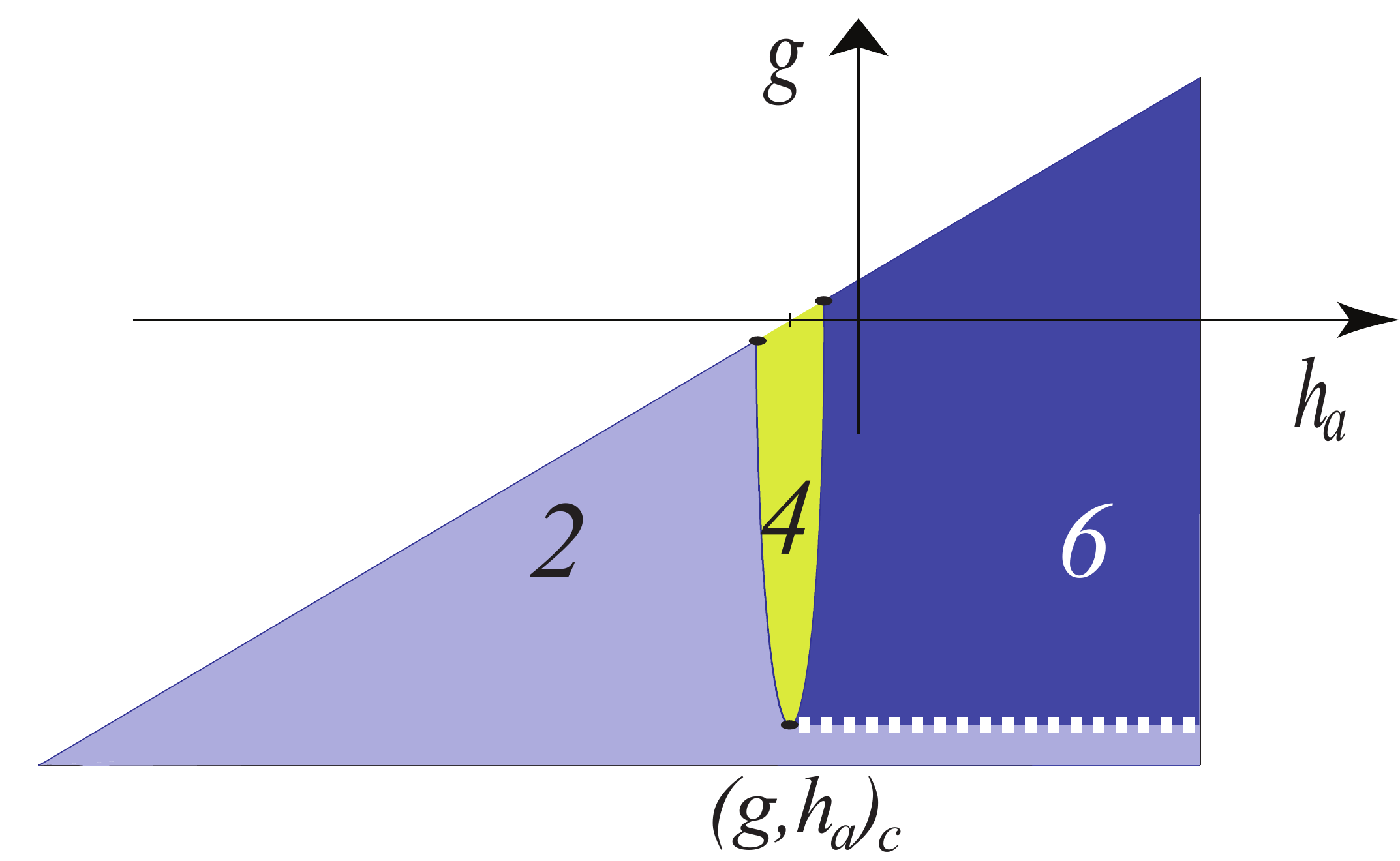}
\caption{(color online) Bifurcation diagram of the parallel state $\mathbf{\mathbf{m}}=\mathbf{e}_{x}$, in the dark zone $\mathbf{\mathbf{m}}=\mathbf{e}_{x}$ is stable. The elliptical-like light zone delimited by $g^2+[h_a-(\beta_x+\beta_z/2)]^2=\beta_z^2/4$ is known as Arnold's tongue. In this region there are 4 equilibria and, the parallel state is unstable. On the left of Arnold's tongue and above the segmented curve $g=-\beta_z/2$, there are 6 equilibria.}
\label{fig-Macrospin}
\end{figure}
This is a Complex Ginzburg-Landau-type equation, which describes the envelope of a nonlinear dissipative oscillator.

An advantage of the Stereographic representation is to guarantee the magnetization normalization and to consider the appropriate degrees of freedom. 
Notice that the switching dynamic  between parallel and anti-parallel state is not well described, since the antiparallel 
state is represented by infinity \cite{Lakshmanan}. This kind of dynamics is not considered in the present work.
To grasp the dynamical behaviour  exhibited by the previous model, 
let us consider that
 the complex amplitude is small, and that the parameters $\alpha, \beta_z/2$ are also small. Introducing 
the renormalized amplitude $ A(\mathbf{r},t)=\psi (\mathbf{r},t) e^{i\pi /4}\sqrt{2\beta_x+\beta_z}$,
after straightforward calculations, Eq.~(\ref{Eq-LLGSterographic}) is approximated at dominant order by
\begin{gather}
\frac{dA}{dt}=-i\nu A-i\left|  A\right|^{2} A-\mu A+\gamma \bar{A},
\label{E-PDNLS}
\end{gather}
where $\mu\equiv-g-\alpha\nu$, $\nu \equiv-h_a-(\beta_x+\beta_z/2)$,  and $\gamma\equiv \beta_z/2$. 
Thus under the above assumptions the nanopillar resonator is described  by Eq.~(\ref{E-PDNLS}), which is known as Parametrically Driven damped NonLinear Schr\"odinger (PDNLS) equation without space. This model has been used to describe parametric resonators~\cite{Landau}.
 
The coefficient $\gamma$ is the intensity of the forcing in usual parametric systems. For instance, it is proportional to the amplitude of the oscillation in vibrated media or the intensity of time-dependent external fields. In the case of the nanopillar $\gamma=\beta_z/2$ is not a control parameter. In the context of the PDNLS amplitude equation $\gamma$ breaks the phase invariance, ie. $A\nrightarrow Ae^{i\phi_0}$. A change of variables of the form $A=Be^{i\omega t}$ (rotating frame)  permits to restore the explicit time-dependent forcing,

\begin{equation}
\frac{dB}{dt}=-i(\nu+\omega) B-i\left|  B\right|^{2} B-\mu B+\gamma e^{-2i\omega t} \bar{B}.
\end{equation} 

Moreover, in this representation the parametric nature of the PDNLS equation is evident.
The parameter $\mu>0$ accounts for dissipation in parametric systems and it models radiation, viscosity and friction, depending on the particular physical context. In our case, this dissipation is the combination of the Gilbert damping and the spin-polarized current. 
Finally, the detuning $\nu$ accounts for the deviation from a half of the forcing frequency. In the case of the nanopillar, $\nu$ is controlled by the external field.

To obtain Eq.~(\ref{E-PDNLS}) we have assumed that $\alpha, \beta_z/2\ll 1$ and that 
 the amplitude  
is a slowly-varying amplitude ($|A|\ll 1$), 
it is, we have the scaling $|A|^2\sim\nu\sim\mu\sim\gamma\sim\partial_t \ll 1$. 
Notwithstanding, the model, Eq.~(\ref{E-PDNLS}), is qualitatively valid outside this limit.

The parallel state $A=0$ is always a solution of Eq.~(\ref{E-PDNLS}).
Decomposing the amplitude into its real and imaginary parts $A(t)=u(t)+i v(t)$ and 
 linearizing around them, we have
\begin{equation}
\frac{d}{dt}
\begin{pmatrix}
u\\
v
\end{pmatrix}
=
\begin{bmatrix}
\gamma-\mu & \nu\\
-\nu & -(\gamma+\mu)
\end{bmatrix}%
\begin{pmatrix}
u\\
v%
\end{pmatrix}.
\end{equation}
Imposing a solution of the form $(u,v)\sim e^{\lambda_\pm t}(u_0,v_0)$, we obtain 
the growth rate relation
$\lambda_\pm=-\mu\pm\sqrt{\gamma^2-\nu^2}$. The stability condition, which corresponds to $\Re e(\lambda_\pm) <0$ is shown in dark areas  in Fig.~\ref{fig-Macrospin}. The elliptical-like light zone of Fig.~\ref{fig-Macrospin} is known as Arnold's tongue in the context of parametric systems, and it accounts for the destabilization of the parallel state for $\mu^2+\nu^2=\gamma^2$. The exact curve of the Arnold's tongue in terms of the original parameters can be obtain from the LLG equation without neglecting $\alpha$, it is $g^2+[h_a-(\beta_x+\beta_z/2)]^2=\beta_z^2/4$. Inside the Arnold's tongue this
model has also the  equilibria  
\begin{equation}
 A_\pm= \pm\left (1- i\sqrt{\frac{\gamma-\mu}{\gamma+\mu}}\right)\sqrt{\frac{\gamma+\mu}{2\gamma}\left(\sqrt{\gamma^2-\mu^2}-\nu\right)}.
\end{equation}
In this region there are four equilibria (see Fig.~\ref{fig-Macrospin}), they are the parallel state $A=0$ (equivalently $\mathbf{\mathbf{m}}=\mathbf{e}_{x}$), the antiparallel state ($\mathbf{\mathbf{m}}=-\mathbf{e}_{x}$) and $ A_\pm$. Crossing the curve of the Arnold tongue for  positive detuning $\sqrt{\gamma^2-\mu^2}=\nu>0$, the $A_\pm$ states and $A=0$, collide together through a pitchfork bifurcation. For greater values of the detuning parameter $\nu$, only the parallel and antiparallel states exist.

For negative detuning and $\gamma>\mu$ (above the dashed 
curve in  Fig.~\ref{fig-Macrospin}), and outside Arnold's tongue $\sqrt{\gamma^2-\mu^2}<|\nu|$, 
the $ A_\pm$ states exist and are stable. Since the $A=0$ equilibrium is also 
stable in this region, it is necessary to have other 2 states $A'_\pm$ that separate 
them in the phases space. 
Which have the 
form
\begin{equation}
 A'_\pm= \pm\left (1+ i\sqrt{\frac{\gamma-\mu}{\gamma+\mu}}\right)\sqrt{\frac{\gamma+\mu}{2\gamma}\left(-\sqrt{\gamma^2-\mu^2}-\nu\right)}.
\end{equation}
In this region (the darkened area in Fig.~\ref{fig-Macrospin}), there are 6 equilibria.
Thus the PDNLS equation describes the homogeneous stationary solutions which have been studied the context of the nano-oscillator\cite{Serpico,Li}.

When  $g\leqslant-\alpha\nu$ the coefficient that rules the dissipation becomes negative, and the magnetization oscillates and moves away from the parallel state. This instability known as Andronov-Hopf 
bifurcation \cite{Andronov}. When it does not saturate the 
magnetization switches to the antiparallel state or reaches another stationary equilibrium. 
Precessions or self-oscillations emerge when this instability saturates. 
In the past years, this regime has been extensively studied experimentally and theoretically in the context of the Spin-transfer torque resonator~\cite{ST_EXP,ST_EXPI,Slavin}. 
This instability does not occur in usual parametric systems since the dissipation coefficient is always positive $\mu>0$.

In brief, the nanopillars driven by a spin-transfer torque effect are equivalent to parametric systems, and then they are well described by the paradigmatic model for parametric systems, the PDNLS equation without space. We will see in the next section the predictions of this model for the nanopillar in the case of a variable magnetization.

\section{Generalization to an inhomogeneous magnetization dynamics}
\label{Sec-Inhomogenity}
The macrospin approximation permits to understand several features of the magnetization dynamics driven with spin torque, even so this approximation is not completely valid because in general both the precession and magnetic reversion are inhomogeneous~\cite{Lee}. There are several approaches to study the non-uniform magnetization dynamics, nevertheless we use here a minimal model with a ferromagnetic exchange torque as the dominant space-dependent coupling in order to understand the emergence of a rich spatio-temporal dynamics.

In the case of an inhomogeneous magnetization $\mathbf{m}(\mathbf{r},t)$, which corresponds to a spatial extension of the nano-oscillator, the magnetic energy $E=\mu_0 M^2_s\int \epsilon dx dy$ of the free layer 
is the integral of the following dimensionless density of energy\cite{Serpico}
\begin{eqnarray}
\epsilon=-\mathbf{m}\cdot {\bf h_a}
- \frac{1}{2}\beta_xm_x^2
+ \frac{1}{2}\beta_z m_z^2
+\frac{1}{2}|\nabla \mathbf{m}|^2,
\label{Eq-EnergyComp}
\end{eqnarray}
where $\{x,y\}$ stands for the spatial coordinates of the free layer. The spatial coordinates have been dimensionless $\mathbf{r}\rightarrow l_{\mathrm{ex}} \mathbf{r}$ in terms of the {\em exchange length} $l_{\mathrm{ex}} \equiv\sqrt{2A/(\mu_0 M^2_s)}$ 
where $A$ is the exchange coupling in the ferromagnet. The gradient operator is defined on the plane 
of the film as $\nabla \equiv \mathbf{e}_{x}\partial_x
+ \mathbf{e}_{y}\partial_y$. The $\beta_x$ and $\beta_z$ coefficients account for both the easy axis and 
the demagnetization in the thin film approximation~\cite{Serpico}. In this approximation, 
the contribution of the demagnetization effect to the magnetic energy density is local, 
and the shape of the thin film is taken into account by the  Neumann boundary condition for the magnetization.

The LLG equation and the effective field are
\begin{equation}
\frac{\partial \mathbf{m}}{\partial t}=-\mathbf{m}
\times \mathbf{h}_{eff}
+\alpha\mathbf{m}\times\frac{\partial \mathbf{m}}{\partial t}
+g\;\mathbf{m}\times(\mathbf{m}\times \mathbf{e}_{\mathbf{x}}),
\label{Eq-LLGSpace}
\end{equation}
\begin{equation}
\mathbf{h}_{eff}\equiv -\frac{1}{\mu_0M_s^2}\frac{\delta E}
{\delta\mathbf{m}}=(h_{a}+\beta_{x}m_{x})\mathbf{e}_{x}-\beta_{z}m_{z}\mathbf{e}_{z}+\nabla^{2}\mathbf{\mathbf{m}}.
\label{Eq-EffF2}
\end{equation}

Notice that gradients come from the ferromagnetic exchange energy, and then the spatial derivatives 
must be written in terms of the coordinates that label the sample, even if it rotates. 
Then the equation of the magnetization of the rotating plate in its co-movil frame is Eq.~(\ref{Eq-LLGRotatingPlateS'}) with an extra term for the spatial dependence.

\begin{eqnarray}
\left. \frac{\partial\mathbf{m}}{\partial t}\right|_{S} =-&\mathbf{m}
\times \mathbf{h}_{eff} +\alpha\mathbf{m}\times \left. \frac{\partial \mathbf{m}}{\partial t} \right|_{S}\nonumber \\
&- \alpha  \Omega_0 \mathbf{m}\times (\mathbf{m} \times \mathbf{e}_{x}).
\label{Eq-LLGRotatingPlateS'2}
\end{eqnarray}
Where $\mathbf{h}_{eff}=(h_{a}+\beta_{x}m_{x})\mathbf{e}_{x}-\beta_{z}m_{z}\mathbf{e}_{z}+\nabla^2 \mathbf{m}$ and the $\nabla \equiv \mathbf{e}_{x}\partial_x+ \mathbf{e}_{y}\partial_y$ operator is defined on the co-movil plane spanned by $(\mathbf{e}_x,\mathbf{e}_y)$. Thus the spatial dependence of $\mathbf{m}$ does not change the equivalence between the nanopillar and the rotating magnet presented in Sec.~\ref{Sec-Equivalence}.
Using the same change of variable of Eq.~(\ref{Eq-CofVar}),  the LLG Eq.~(\ref{Eq-LLGSpace}) reads
\begin{figure}[t!]
\includegraphics[width=7.5 cm]{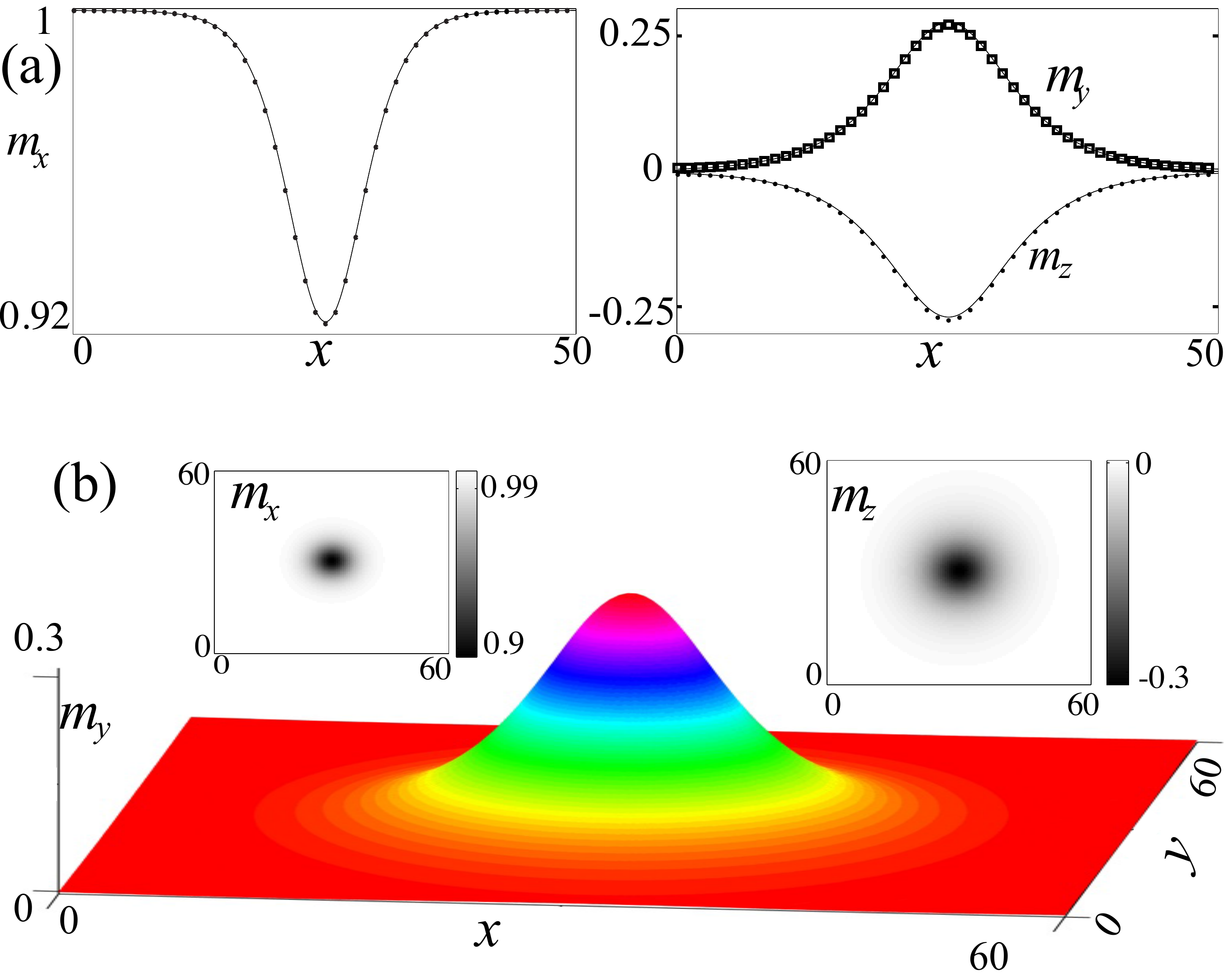}
\caption{(color online) Dissipative solitons  in one- and two-dimensional nanopillars (with a square cross-section) with $\beta_x=0.5$, $\beta_z=1$, and $\alpha=0.05$. 
(a) One-dimensional soliton for $g=-0.4999$, $h_a=-0.97$, points account for the numerical 
integration of the LLG equation, and line accounts for  the analytical solution 
given by Eq.~(\ref{Eq-Soliton}). (b) Soliton in two-dimensions, $g=-0.49995$, $h_a=-0.99$, the three-dimensional plot shows the profile of the component $m_y$, while the insets show the $m_x$ and the $m_z$ components.} 
\label{fig-Solitons}
\end{figure}
\begin{eqnarray}
\left(i+\alpha\right)&\partial_{T}\psi =\left(ig-h_{a}\right)\psi-\frac{\beta_{z}}{2}\left(\psi-\overline{\psi}\right)\frac{1+\psi^{2}}{1+|\psi|^{2}} 
\nonumber \\
&-\beta_{x}\psi\frac{1-|\psi|^{2}}{1+|\psi|^{2}}+\nabla^{2}\psi-2\frac{\overline{\psi}}{1+|\psi|^{2}}\left(\nabla\psi\right)^{2}.
\label{Eq-LLGSterographicComp}
\end{eqnarray}
Which describes the envelope of coupled nonlinear oscillators. 
Due to the complexity of this equation, we will consider a simple limit, which 
permits to grasp its dynamics. Using the small amplitude that varies 
slowly in space $A(\mathbf{r},t)=\psi (\mathbf{r},t) e^{i\pi /4}\sqrt{2\beta_x+\beta_z}$, we obtain
\begin{gather}
\partial_{\tau}A=-i\nu A-i\left|  A\right|^{2} A-i \nabla^{2}A-\mu A+\gamma \bar{A},
\label{E-PDNLSComp}
\end{gather}
which is the PDNLS model. The extra term with spatial derivatives describes dispersion.

\subsection{Parametric textures for nanopillars}
\label{Sec-Textures}

The above model, Eq.~(\ref{E-PDNLSComp}) has been extensively used to study 
the pattern formation, particularly, this model exhibits solitons, oscillons, periodic textures, 
complex behaviours, among others.  To verify these predictions, 
we compare with the numerical solutions of Eq.~(\ref{Eq-LLG}) in two geometrical configurations. 
The first is a one-dimensional free layer, that is, a nanopillar for which $\mathbf{m}(\mathbf{r},t)\approx\mathbf{m}(x,t)$ 
and the second is a two-dimensional nanopillar with a square cross-section. Different transversal 
lengths are used in simulations, all of them displaying the same qualitative aspects of 
the solutions. The simulations are conducted using a fifth order 
Runge-Kutta algorithm with constant step-size for time integration 
and finite differences for spatial discretization. The spatial differential operators are approximated 
with centered schemes of order-6 and specular (Neumann) boundary conditions are used. 

\subsubsection{Dissipative solitons}

Analytical solutions for the dissipative soliton are known in one dimension\cite{ClercCoulibalyLaroze2008,Barashenkov,Zarate11}. 
In two-dimensions dissipative solitons are observed, however without analytical expressions.
 From this result and using the stereographic change of variable, we find the following analytical form for 
 magnetic dissipative solitons in one dimension
 \begin{eqnarray}
m_{x} & = & \frac{2\beta_{x}+\beta_{z}-R(x)^2}{2\beta_{x}
+\beta_{z}+R(x)^2}, \nonumber \\
\left(\begin{array}{c}
m_{y}\\
m_{z}
\end{array}\right) & = & \frac{2R(x)\sqrt{2\beta_{x}+\beta_{z}}}{2\beta_{x}+\beta_{z}+R(x)^2} \left(\begin{array}{c}
\cos\varphi_{0}\\
\sin\varphi_{0}
\end{array}\right),
\label{Eq-Soliton}
\end{eqnarray}
with $ \sin\left(2\varphi_{0}\right) \equiv 2g/\beta_{z}$, 
$R\equiv\sqrt{2\delta}\text{sech}\left[\sqrt{\delta}\left(x-x_{0}\right)\right]$ 
and $\delta  \equiv  h_{a}+\beta_{x}+\beta_{z}/2+\sqrt{\left(\beta_{z}/2\right)^{2}-g^{2}}$. 
The width of the soliton is controlled by the external field. The typical sizes are about $10l_{ex}$.
\begin{figure}[t]
\includegraphics[width=7.6 cm]{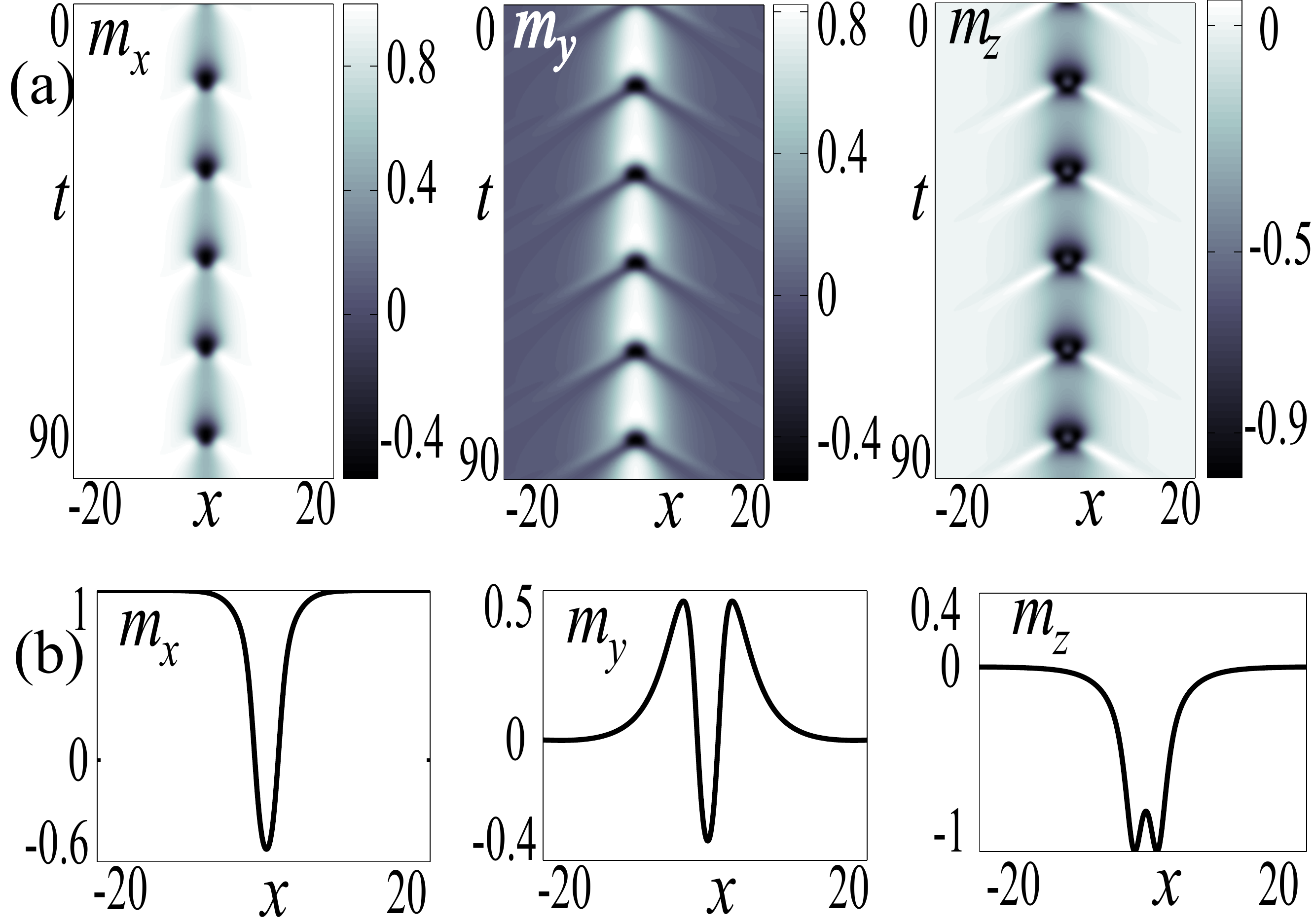}
\caption{(Color online)Breather or Oscillon solution for $g=-0.33$, $h_a=-0.51$. 
(a) is the spatio-temporal diagram. (b) shows the magnetization components at the time for which $m_x$ reaches its minimal position. Typical oscillation periods are about $\Delta t\approx 16(\gamma M_s)^{-1}$, which is about $\Delta t\approx 51ps$ for a $3nm$ thick cobalt free layer.} 
\label{fig-Breathers}
\end{figure}
\begin{figure}[b!]
\includegraphics[width=8.6 cm]{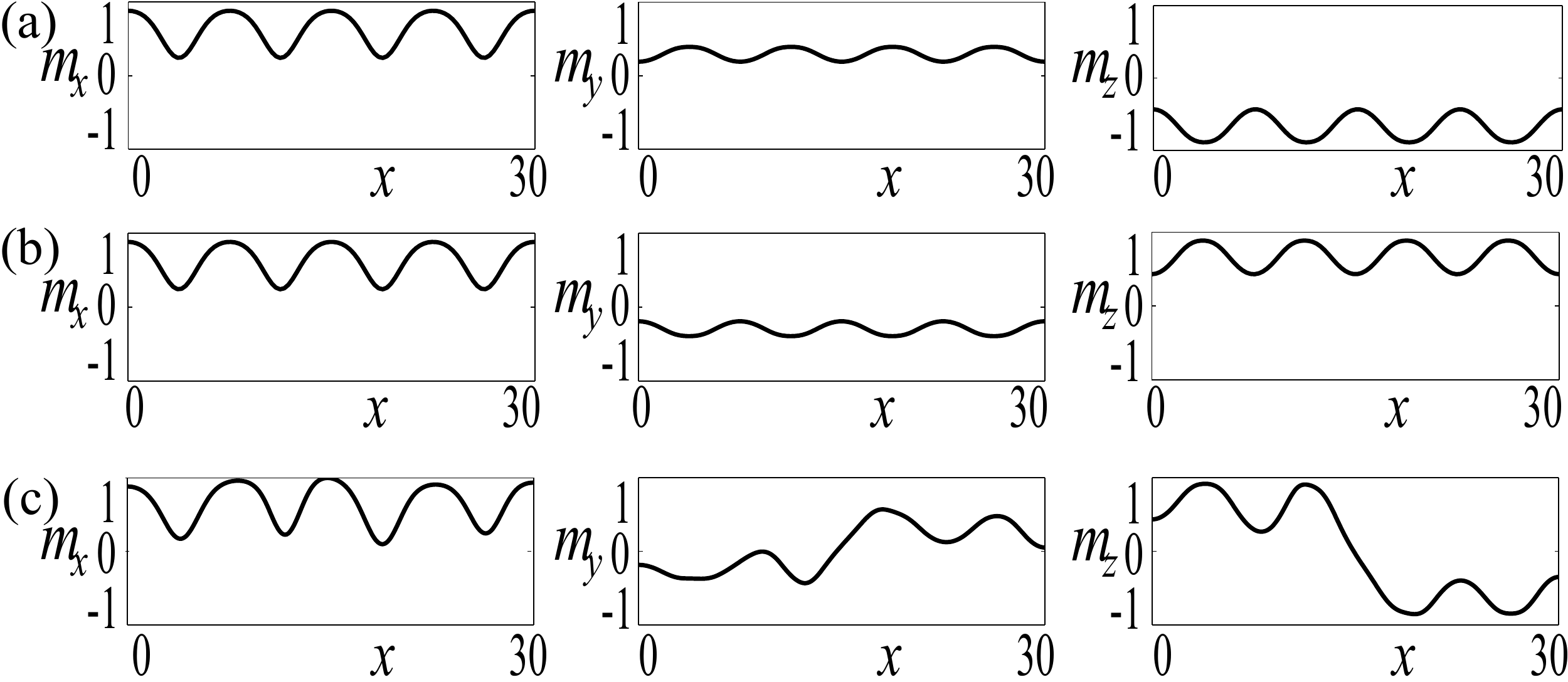}
\caption{Dissipative structures for $g=-0.37$, $h_a=-0.75$, inside Arnold's tongue. 
(a) and (b) represent pattern states. (c) Is a slowly moving kink connecting the (a) and (b) patterns.} 
\label{fig-Kinks}
\end{figure}

\begin{figure}[t!]
\includegraphics[width=8.6 cm]{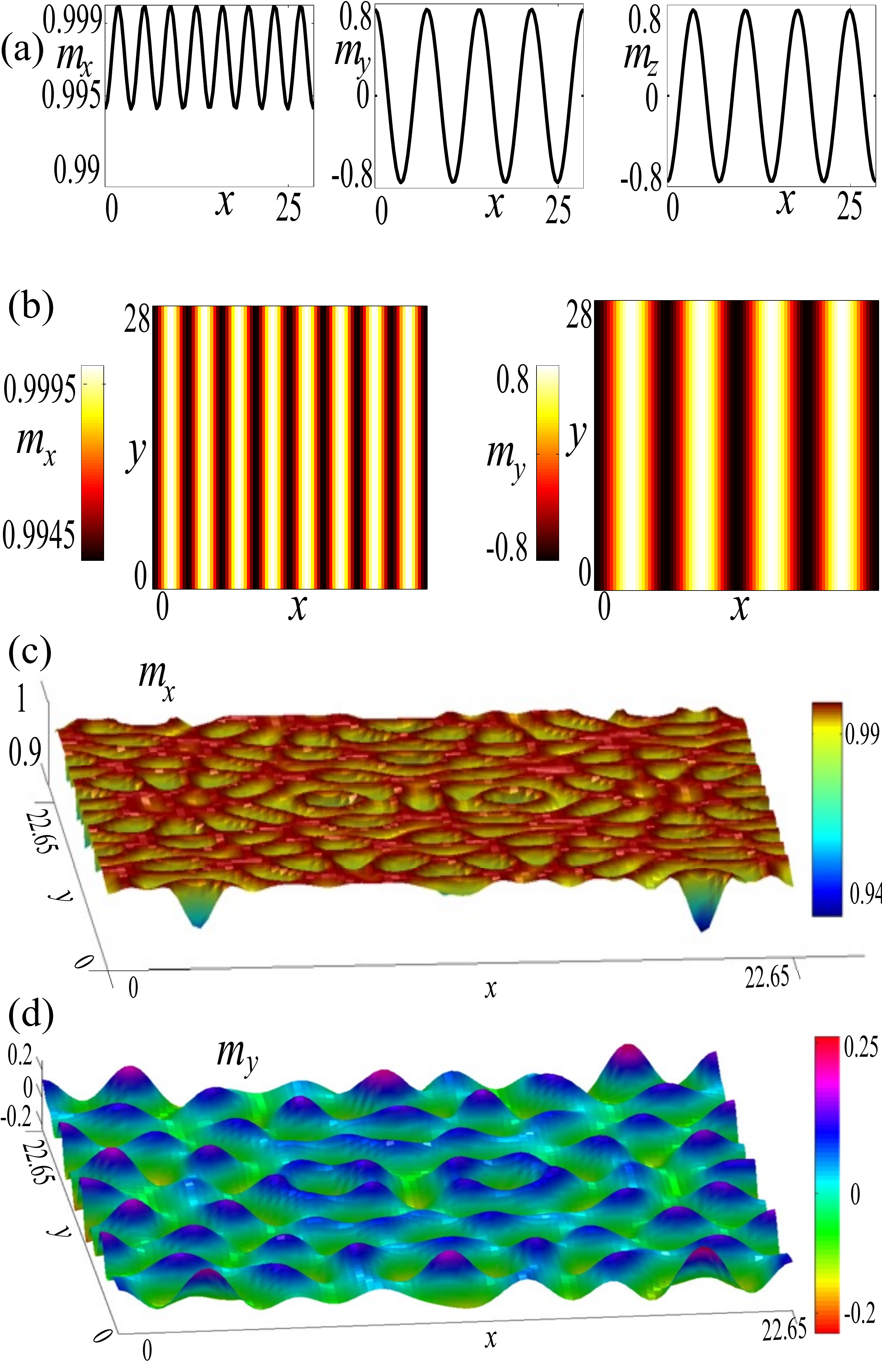}
\caption{(color online) Patterns induced by the Spin-polarized current.
(a) One-dimensional state for $g=-0.49999$, $h_a=-1.8$, as predicted by Eq.~(\ref{Eq-Patterns}),  $m_y\approx -m_z$. Notice the norm conservation implies that $m_x\approx 1-0.5(m_y^2+m_z^2)$ oscillates with a half of the wavelength of the other two components. (b) Bi-dimensional pattern for the same parameters used in (a), the component $m_z$ (not shown) is the negative of $m_y$. (c) and (d) show the magnetization components $m_x$ and $m_y$ for a superlattice pattern obtained with $g=-0.4999$, $h_a=-6$. As for the other patterns,  $m_y\approx -m_z$.} 
\label{fig-Patterns}
\end{figure}
\begin{figure}[t!]
\includegraphics[width=7.0 cm]{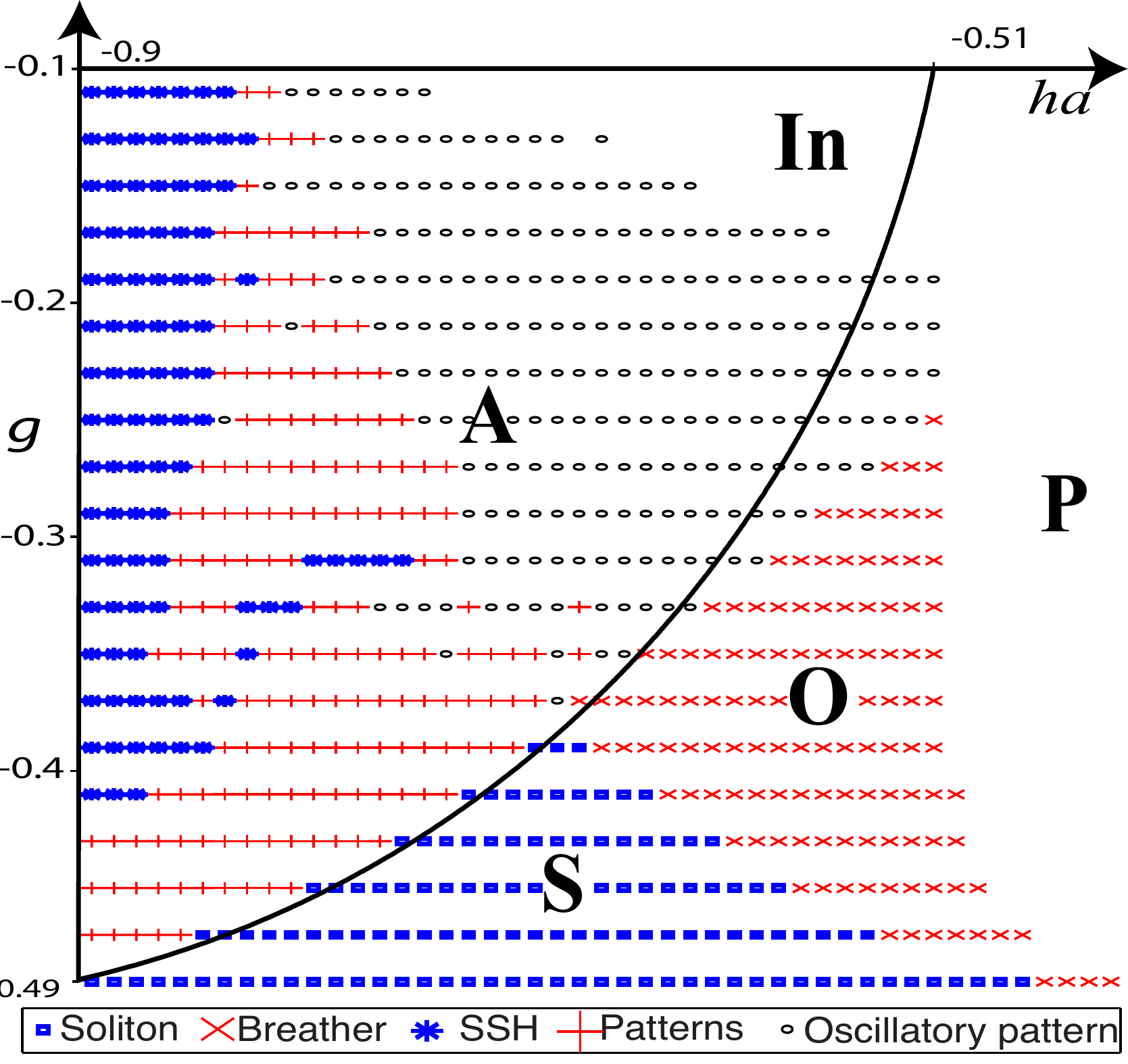}
\caption{(Color online) Phase diagram of LLG model, Eq.~(\ref{Eq-LLG}). The S-region represents solitons region,  
O-region stands for breather solitons (Oscillons) region, A-region is the Arnold's tongue. The In-Regions accounts for  
inhomogeneous dynamical states far from the parallel configuration. In the zone P, only the Parallel state is observed}.
\label{fig-PhaseDiagram}
\end{figure}
Figures~\ref{fig-Solitons}a shows the analytical results compared with numerical 
simulations of the LLG equation, which presents a
quite good agreement for small amplitude 
solitons, i.e. for $\delta\ll1$. Furthermore,
figure~\ref{fig-Solitons}b illustrates the dissipative solitons 
observed numerically in two dimensions.
We note that these solitons are well described by hyperbolic secant function, 
which have been obtained using variational methods \cite{Anderson}.

Dissipative solitons are observed in the region of parameter space bounded 
by   $\beta_z^2/2-(|h_a|-(\beta_x+\beta_z/2))^2=g^2$ and $\beta_z/2=|g|$. 
This region is analytically inferred from the amplitude Eq.~(\ref{E-PDNLSComp}). 
Figure~\ref{fig-PhaseDiagram} shows the respective phase diagram 
of the LLG equation, the region of dissipative solitons is denoted by S-region.

Increasing the difference between injection and dissipation, $\gamma-\mu$, dissipative 
solitons undergo an Andronov-Hopf bifurcation, 
generating oscillatory localized states or breather solitons  characterized by exhibiting  shoulders in the
 amplitude profile \cite{Barashenkov2011}.
Figure~\ref{fig-Breathers} illustrates this kind of solution. Similar solutions have also been reported
in a magnetic wire forced by a transversal uniform and oscillatory magnetic field \cite{Urzagasti},
which corresponds to a parametric system. These oscillatory solutions are observed in O-region 
of the bifurcation diagram shown in Fig.~\ref{fig-PhaseDiagram}. Notice that, for spin-transfer torques that favour the parallel state, 
the nanopillar can also behave as a nano-oscillator.
 
\subsubsection{Pattern states}

Let us introduce $A$-region of the bifurcation diagram (cf. Fig.~\ref{fig-PhaseDiagram}),
which is circumscribed by the curve $\beta_z^2/2-(|h_a|-(\beta_x+\beta_z/2))^2=g^2$, 
in the Arnold's tongue.
Inside this region the quiescent state $A = 0$ is unstable, giving rise to a non zero uniform state,  
stationary and oscillatory patterns. Figures \ref{fig-Kinks}a and \ref{fig-Kinks}b 
show stable stationary patterns that exist inside the Arnold's tongue, 
Fig. \ref{fig-Kinks}c shows a propagative wall that connects the patterns. 
In addition, the  PDNLS model is characterized by exhibiting supercritical patterns at $\gamma=\mu$ ($\beta_z/2=|g|$), 
growing  with power law $1/4$ 
as a function of the bifurcation parameter \cite{Coullet}. Recently, such dissipative structures
induced by spin-transfer torques in nanopillars have been characterized numerically and theoretically~\cite{LeonClercCoulibaly}, where the spatial textures emerge from a spatial supercritical quintic bifurcation.
In one spatial dimension, the magnetic patterns read at dominant order by
\begin{equation}
\binom{m_{y}}{m_{z}}\approx2\left[  \frac{4\beta_{z}\left(  g-g_{c}\right)
}{(6\beta_{x}+3\beta_{z}-2k_{c}^{2})^{2}}\right]  ^{1/4}\binom
{\cos(k_{c}x)}{-\cos(k_{c}x)},\label{Eq-Patterns}
\end{equation}
 
and $m_x\approx 1-(m_y^2+m_z^2)/2$. Figure~\ref{fig-Patterns} shows a pattern solution. The wavelenght of the periodic structures, $2\pi/k_c=2\pi/\sqrt{-h_a-\beta_x-\beta_z}$, is controlled by the external field $h_a<0$.
In two spatial dimensions the system shows the emergence of stripe patterns or superlattices at the onset of bifurcation \cite{LeonClercCoulibaly}. The phases diagram of the textures is controlled by a single parameter that 
accounts for the competition between the external magnetic
field, anisotropy, exchange, and the critical spin-polarized
current.  When the anisotropy is dominant over the external field 
the system exhibits striped patterns (Fig.~\ref{fig-Patterns}b), however, when
the external field drives the dynamics, the system
presents superlattice (Fig.~\ref{fig-Patterns}c and Fig.~\ref{fig-Patterns}d) as stable equilibria. Indeed, external fields pointing against the near parallel states
favor the formation of more sophisticated spatial textures.
Since the electric resistance $R[\mathbf{M}_0\cdot\mathbf{m}]$ of the nanopillar depends on the relative orientation~\cite{Lee} of the fixed $\mathbf{M}_0$ and free $\mathbf{m}$ layers, and $\mathbf{M}_0\cdot\mathbf{m}=m_x\approx 1-(m_y^2+m_z^2)/2$ the signature of the patterns is a time-independent resistance that increases a square root of the current $R=R_p+\eta\cdot(g-g_c)^{1/2}$ when $g$ is negative and goes to zero.  The parameter $\eta$ contains all the information of the applied field, anisotropies, and geometry.

Notice that according to the PDNLS model, Eq.~(\ref{E-PDNLS}) and Eq.~(\ref{E-PDNLSComp}), the parametric resonance occurs when $\nu\approx 0$ and $\gamma\approx \mu$, or equivalently $(g,h_a)_c=-(\beta_z/2,\beta_x+\beta_z/2)$. 
 For a $3nm$ thick material with saturation magnetization similar to cobalt~\cite{Serpico}, it is $M_s\simeq 1.4$ $10^6 A/m$,   the critical current density is $J_c=J(g_c)\approx-\beta_z\cdot 10^9Acm^{-2}$ for a constant $\mathcal{P}(m_x)\approx 1$ polarization function. Since localized states and patterns appear for currents that are fractions of the critical current $|J|\sim 3|J_c|/5$, the smaller the $\beta_z$ parameters is, the smaller the spin-polarized currents required to observe the parametric phenomenology. Most of our results use $\beta_x=1/2$ and $\beta_z=1$, nevertheless we have conducted numerical simulations for different values of $\beta_z$ for $\beta_x$ in order to achieve the parametric resonance at arbitrary small currents, and the predictions of Eq.~(\ref{E-PDNLS}) and Eq.~(\ref{E-PDNLSComp}) remain unchanged. The robustness of this parametric phenomenology is a characteristic of systems near their parametric resonance. 
\section{Conclusions and remarks} 
\label{Sec-Conclusions}
We have shown that nanopillars under the effect of a direct electric current are equivalent to
simple rotating magnetic plates. 
The latter system is characterized by displaying a parametric instability. 
This equivalence permits to transfer the known results of the self-organization of parametric systems to the magnetization dynamics induced by the spin-transfer torque effect. In particular, we have shown that for spin-polarized currents that favor the parallel state the system is governed by the PNDLS equation, and then the magnetization exhibits localized states and patterns both in one and two spatial dimensions.
Numerical simulations show a quite good agreement with the analytical predictions.

\bigskip
The authors thank E. Vidal-Henriquez the critical reading of the manuscript. M.G.C. thanks the financial support of FONDECYT project
1120320. A.O.L. thanks Conicyt fellowship {\it Beca Nacional}, Contract number
21120878.

\end{document}